\documentclass[a4paper]{revtex4}
\usepackage[top=1in, bottom=1in, left=1in, right=1in]{geometry}
\usepackage{amsmath}
\usepackage{amssymb}
\usepackage{amsfonts}
\usepackage{graphicx,bm}
\usepackage{dcolumn}
\usepackage{epsfig}

\def \lleq {\lower0.9ex\hbox{ $\buildrel < \over \sim$} ~}
\def \ggeq {\lower0.9ex\hbox{ $\buildrel > \over \sim$} ~}

\def \beq  {\begin{equation}}
\def \eeq  {\end{equation}}
\def \ber  {\begin{eqnarray}}
\def \eer  {\end{eqnarray}}

\newcommand{\be}{\begin{equation}}
\newcommand{\ee}{\end{equation}}
\newcommand{\ba}{\begin{eqnarray}}
\newcommand{\ea}{\end{eqnarray}}
\newcommand{\bea}{\begin{eqnarray*}}
\newcommand{\eea}{\end{eqnarray*}}

\begin{document}
\title{Dynamics of interacting quintessence}

\author{M. Shahalam$^1$\footnote{E-mail address: mohdshahamu@gmail.com},
S. D. Pathak$^2$\footnote{E-mail address: prince.pathak19@gmail.com},
M. M. Verma$^2$\footnote{E-mail address: sunilmmv@yahoo.com},
M. Yu. Khlopov$^{3,4}$\footnote{E-mail address: khlopov@apc.univ-paris7.fr},
R. Myrzakulov$^5$\footnote{E-mail address: rmyrzakulov@gmail.com}}

\affiliation{$^1$Center For Theoretical Physics, Jamia Millia Islamia, New Delhi 110025, India\\
$^2$Department of Physics, University of Lucknow, Lucknow 226 007, India\\
$^3$National Research Nuclear University ``MEPHI" (Moscow Engineering Physics Institute), 115409 Moscow, Russia \\
$^4$APC laboratory 10, rue Alice Domon et L\'eonie Duquet 75205 Paris Cedex 13, France\\
$^5$Department of General and Theoretical Physics, Eurasian National University, Astana, Kazakhstan}

\begin{abstract}
 In this paper, we investigate coupled quintessence with scaling potential assuming specific forms of the coupling as $A$ namely, $\alpha \dot{\rho_m}$, $\beta \dot{\rho_{\phi}}$ and $\sigma (\dot{\rho_m}+\dot{\rho_{\phi}})$, and present phase space analysis for three different interacting  models. We focus on the attractor solutions that can give rise to late time acceleration with $\Omega_{DE}/\Omega_{DM}$ of order unity in order to alleviate the coincidence problem.
 
\end{abstract}
\maketitle

\section{Introduction}
\label{intro}
A large number of cosmological observations \cite{planck, perlmutter, riess, Spergel, Komatsu} reveal  that  our Universe is experiencing an accelerated expansion at present  and the  transition from deceleration phase to acceleration phase took place in the recent past \cite{riess}. In the standard Einstein gravity, the late time cosmic acceleration is driven by an exotic energy component with huge negative pressure filling the Universe, known as `{\it dark energy}' \cite{review}. One of the simplest candidate of dark energy (DE) is the cosmological constant (CC) $\Lambda$. However, it is plagued with difficult theoretical issues such as fine tuning and cosmic coincidence problem \cite{derev}. This is important to explore whether dark energy is cosmological constant or it has dynamics. To this effect, a variety of dynamical dark energy models have been explored in the references \cite{ratra, caldwell, amna, alam, alam2, sd, quint, phant,Boisseau:2000pr, ArmendarizPicon:2000ah, quintom, saibal, ira}. The models of unstable dark matter with a non zero cosmological constant can also mimic such dynamics \cite{udm}. Alternatively, large scale modification of gravity has been used to obtain late time cosmic acceleration. At present late time cosmic acceleration is treated as an established phenomenon however its underlying cause is still unknown. Within the framework of Einstein gravity and modified theories of gravity, numerous models can explain the said phenomenon.

Although $\Lambda$CDM model is consistent with present observations, yet there is no satisfactory argument for coincidence problem. Interaction of dark energy with dark matter is one novel approach that might address the mentioned problem. The interacting dark energy models have been recently proposed by several authors \cite{w1,x1,x4,x5,a4,q1}. The interaction between dark energy and dark matter may enhance the dark matter, and also affect structure formation. The investigation of phase space analysis is the one conclusive test for dark energy models. Specifically, the attractor solutions are independent for a wide range of initial conditions. If the dark energy models have $\Omega_{DE}/ \Omega_{DM}$ of the order 1 and an accelerated scaling attractor solution, then the coincidence problem can be alleviated. The non-interacting quintessence \cite{expon, Copeland} and quintom \cite{hwe} models show late time accelerated attractors, and possess $\Omega_{DE}$ as 1, therefore, they do not provide an adequate solution for coincidence problem. In the literature, two forms of interactions have been discussed namely local and non- local. Local forms of interactions are directly proportional to energy density whereas non- local forms are directly proportional to Hubble parameter $H$ and energy density $\rho$. In this paper we consider local forms of interactions proportional to energy density. Some of the local forms   have been discussed in references \cite{Boehmer, Cen, Malik, Zia}. Note that some of the choices of interacting terms appeared implicitly in the literature \cite{a35}. There is also approach to discuss the interacting term without the assumption of a specific form of interacting term \cite{a36}. The plan of the work is organised as follows: In section \ref{QC} we establish the interacting quintessence cosmological framework and construct an autonomous dynamical system which is worthy for phase space investigation. In section \ref{PSA} we discuss phase space analysis and find stationary points and their stability for three interacting quintessence cosmological models. Our results are presented in section \ref{conc}.

\section{Quintessence cosmology}
\label{QC}
We consider two components first one is canonical scalar field (quintessence) as a source of dark energy in spatially flat Universe, and second one is matter (Baryonic+DM). The total energy density of the Universe is conserved, and the individual components of energy density may not be conserved. Thus, we are considering following conservation equations of energy density as:
\begin{eqnarray} 
\dot{\rho_{tot}}+3H(1+w_{tot})\rho_{tot}=0,    \nonumber \\    
\dot{\rho_{\phi}}+3H(1+w_{\phi})\rho_{\phi}=-A,    \nonumber \\     \dot{\rho_{m}}+3H(1+w_{m})\rho_{m}=A,
\label{conser}
\end{eqnarray}
where $\rho_{tot}=\rho_{\phi}+\rho_{m}$, 
$1 + w_{\phi}=\dot{\phi}^{2}/\rho_{\phi}$,  $w_{m}=0$ is the equation of state of matter, $A$ is the interaction strength and $H$ is the Hubble parameter which is given as
\begin{eqnarray}
H^{2}=\frac{8\pi G}{3} \rho_{tot}
\label{hub}
\end{eqnarray}
The sign of $A$ gives information about the direction of flow of energy between two components. There are 3 cases:\\
Case I: If $A \ > 0$, in this case transfer of energy occurs from  quintessence to dark matter. Consequently, quintessence losses self strength and gives dark matter.\\
Case II: If $A \ < 0$, under this condition dark matter losses its strength and there is energy transfer from  dark matter to quintessence.\\
Case III: If $A = 0$, under this condition quintessence do not interact with dark matter, and no energy transfer at all between two components considered in the literature. Therefore, we are not considering this case.

Since, there is no fundamental theory of dark energy and dark matter interaction (interaction in dark sector) at present, therefore it is not possible to construct the functional form of interaction strength $A$ from first principle. Different forms of interaction strength (linear and non-linear) have been considered by several authors \cite{y1, tapan, z1,z2,z3, chena, saridakisa, Maartensa,nicola}. Motivated from the left hand side of the energy conservation equation (\ref{conser}) it is natural that $A$ should be the function of Hubble parameter and energy density that is
\begin{eqnarray}
A= A(H, \rho_m, \rho_{\phi})
\label{A0}
\end{eqnarray}
Here we consider three specific forms of interaction strength heuristically as:
\begin{eqnarray}
\label{A1}
A&=& \alpha\dot{\rho_{m}},\\
\label{A2}
A&=& \beta\dot{\rho_{\phi}},\\
\label{A3}
A&=& \sigma (\dot{\rho_{m}}+\dot{\rho_{\phi}})
\end{eqnarray}
Since, Hubble parameter has the dimension of inverse of time, and inverse of time is sitting in the rate change of energy density, see equations (\ref{A1})$-$ (\ref{A3}), therefore we are not incorporating $H$ separately in the functional form of $A$. The evolution equations in a flat Friedmann- Lemaitre- Robertson- Walker (FLRW) Universe can be written as:
\begin{eqnarray}
\label{hdot}
H^{2}&=&\frac{\kappa^2}{3} (\rho_{m}+\rho_{\phi})\\ \nonumber
2H \dot{H}&=&\frac{\kappa^2}{3} (\dot{\rho_{m}}+\dot{\rho_{\phi}})
\end{eqnarray}
where $\kappa^2= 8\pi G$, $\rho_{\phi}=\frac{1}{2}\dot{\phi^2}+ V(\phi)$ and 
$p_{\phi}=\frac{1}{2}\dot{\phi^2}- V(\phi)$.
We introduce following dimensionless parameters
\begin{eqnarray}
X^{2}=\frac{\kappa^{2}\dot{\phi}^{2}}{6H^{2}}; \quad Y^{2}=\frac{\kappa^{2}V}{3H^{2}}; \quad
\lambda =-\frac{V'}{\kappa V}
\end{eqnarray}
to form an autonomous system of evolution equations (\ref{conser}) and (\ref{hdot}) as:
\begin{eqnarray}
\label{n15}
\dfrac{dX}{dN}&=&-3X + \sqrt{\frac{3}{2}}\lambda Y^{2} - X\frac{\dot{H}}{H^{2}}\\ \nonumber
\dfrac{dY}{dN}&=&-\sqrt{\frac{3}{2}}\lambda XY - Y\frac{\dot{H}}{H^{2}}
\end{eqnarray}
where $N=\ln a$. The total equation of state and field density parameter are given as:
\begin{eqnarray}
\label{wtot}
W_{tot}&=& -1 - \frac{2\dot{H}}{3H^2}\\ \nonumber
\Omega_{\phi}&=&\frac{\kappa^2 \rho_{\phi}}{3H^2}= X^2+Y^2
\end{eqnarray}
The condition for acceleration is $W_{tot} < -\frac{1}{3}$.
\section{Phase space analysis: Stationary points and their stability}
\label{PSA}
In this section we shall use an autonomous system [equation (\ref{n15})], which is appropriate for obtaining stationary points and stability. The stationary points shall be obtained by equating the left hand side of equation (\ref{n15}) to zero. Their stability will be confirmed from the sign of the corresponding eigenvalues, which will be obtained numerically.

\subsection{Interacting model I}
\label{M1}
In this model we consider the following specific form of interaction strength
\begin{equation}
A=\alpha\dot{\rho_{m}}
\end{equation}
Using equation (\ref{hdot}) for this phenomenological form of interaction we have
\begin{equation} \frac{\dot{H}}{H^2}=-\frac{3}{2}\left[\frac{1-Y^{2}+X^{2}(1-2\alpha)}{1-\alpha} \right]\label{n16}\end{equation}
By using $\Omega_{\phi}+\Omega_{m}=1$ where $\Omega_{m}=1-X^{2}-Y^{2}$. Thus, from equation (\ref{n15}) with equation (\ref{n16}) we have following form of autonomous system
\begin{eqnarray}
\label{n18}
\dfrac{dX}{dN}&=&-3X + \sqrt{\frac{3}{2}}\lambda Y^{2} +\frac{3}{2}X\left[\frac{1-Y^{2}+X^{2}(1-2\alpha)}{1-\alpha} \right]\\ \nonumber
\dfrac{dY}{dN}&=&-\sqrt{\frac{3}{2}}\lambda XY + \frac{3}{2}Y\left[\frac{1-Y^{2}+X^{2}(1-2\alpha)}{1-\alpha} \right]
\end{eqnarray}
Using equations (\ref{wtot}) and (\ref{n16}), the total equation of state for this model can be written as
\begin{eqnarray}
\label{wtot1}
W_{tot}&=&-1 + \left[\frac{1-Y^{2}+X^{2}(1-2\alpha)}{1-\alpha} \right]\equiv W_{\phi} \Omega_{\phi}
\end{eqnarray}
\begin{figure*}
\begin{center}
$\begin{array}{c@{\hspace{1in}}c}
\epsfxsize=2.4in
\epsffile{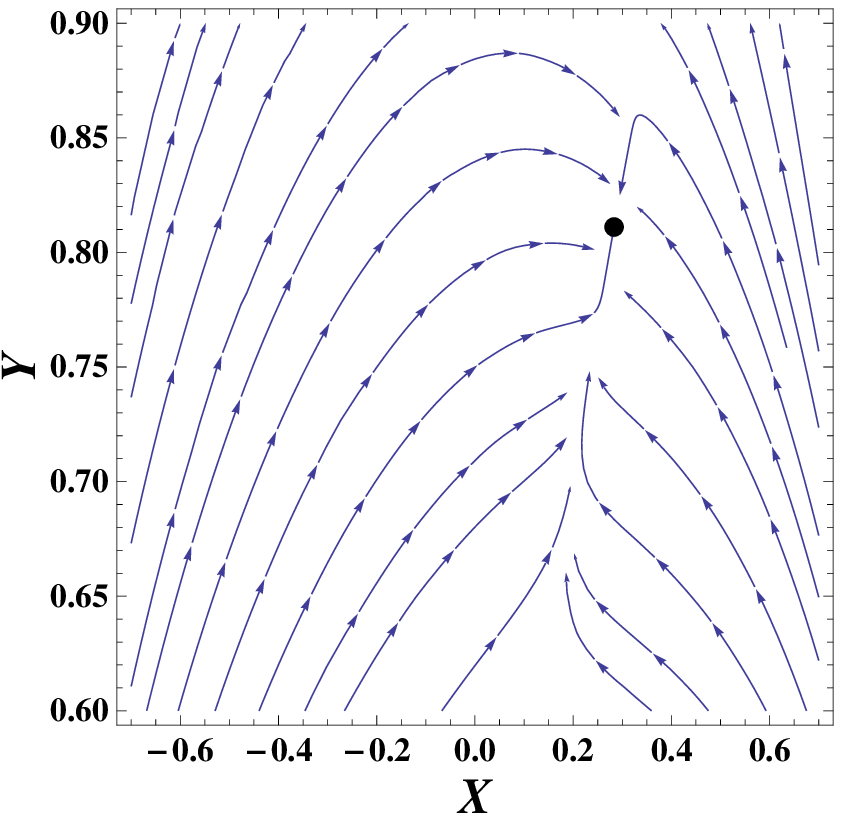} &
	\epsfxsize=2.4in
	\epsffile{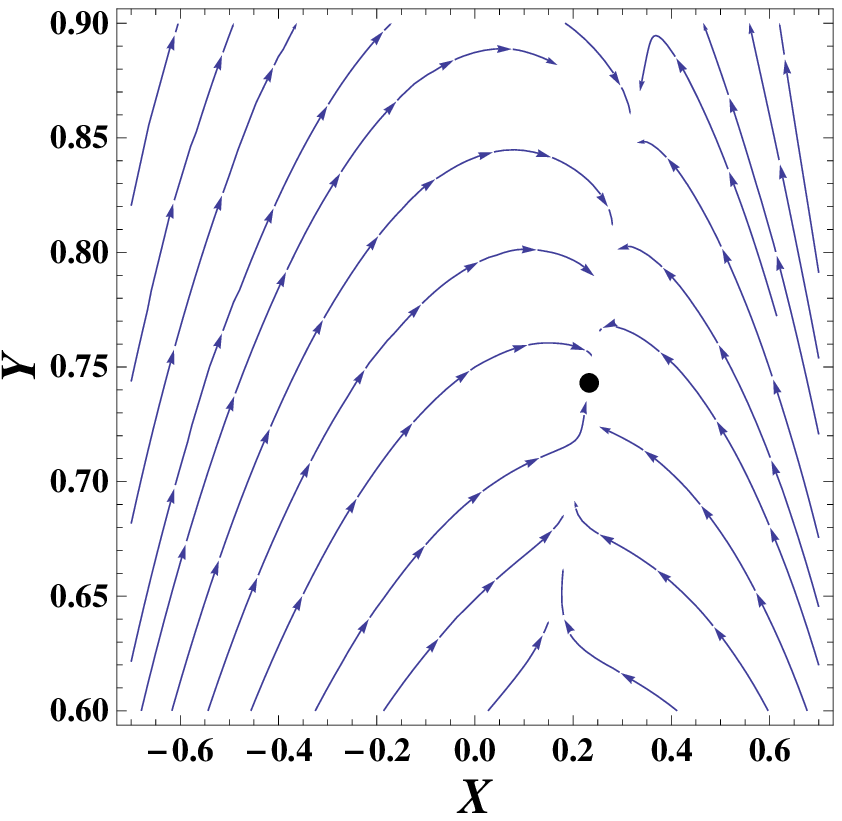} \\ [0.4cm]
\mbox{\bf (a)} & \mbox{\bf (b)}
\end{array}$
$\begin{array}{c@{\hspace{1in}}c}
\epsfxsize=2.4in
\epsffile{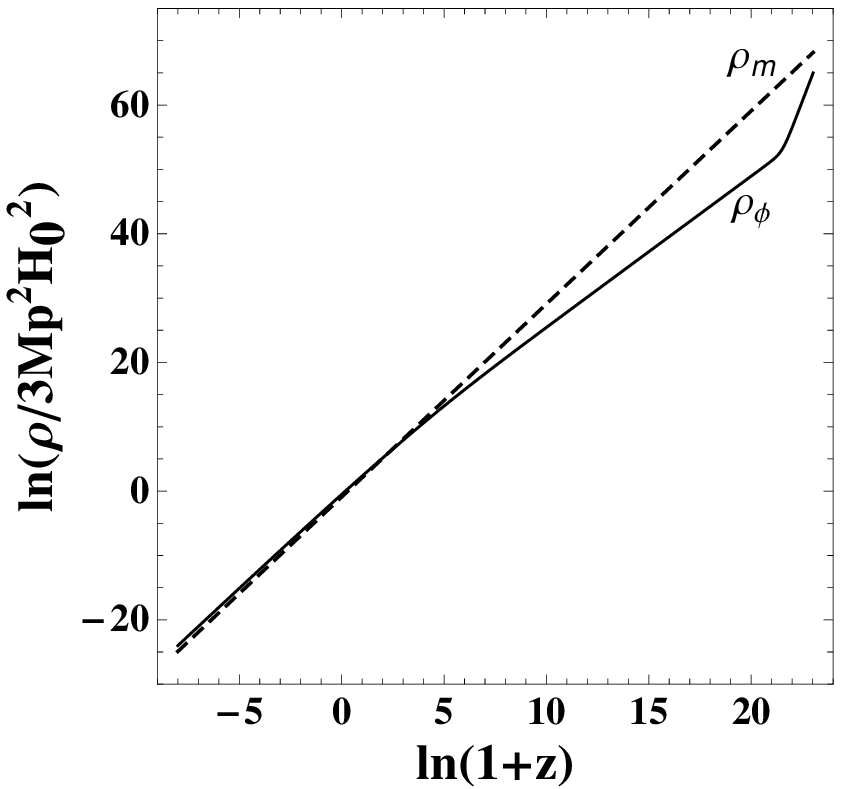} &
	\epsfxsize=2.4in
	\epsffile{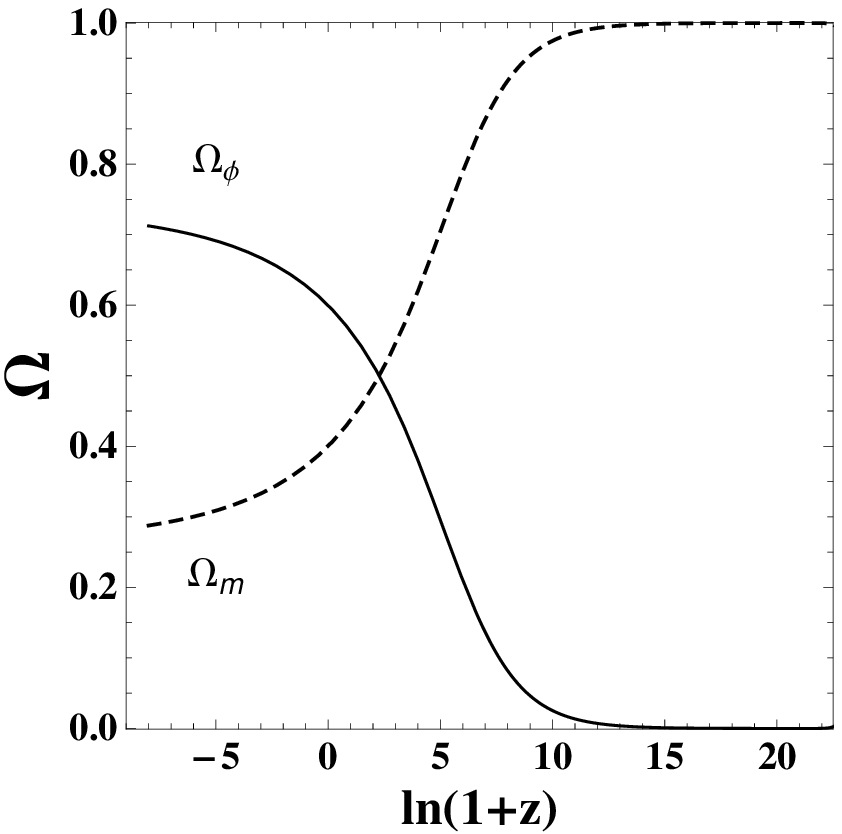} \\ [0.4cm]
\mbox{\bf (c)} & \mbox{\bf (d)}
\end{array}$
\end{center}
\caption{ \small This figure shows the phase portrait, evolution of energy density and density parameter for model I for the stable fixed point 5 which is an attractive node. The panel (a) corresponds to $\alpha= -3.6$ and $\lambda= 0.94$, for these values of the parameters we obtain $\Omega_{\phi}=0.73$, $W_{tot}=-0.78$, $W_{\phi}=-1.06$ and an accelerating attractor solution. The panel (b) corresponds to $\alpha= -4.6$ and $\lambda= 0.94$, correspondingly we get $\Omega_{\phi}=0.60$, $W_{tot}=-0.82$, $W_{\phi}=-1.36$ and an accelerated attractor solution. In both the panels, black dots represent attractor stable point. The panels (c) and (d) have same values of the parameters as panel (a), and show scaling behaviours that provide an accelerated expansion.}
\label{figint1}
\end{figure*}
The critical points of autonomous system (\ref{n18}) could be obtained by setting $\dfrac{dX}{dN}=0$ and $\dfrac{dY}{dN}=0$ simultaneously. Thus, we have following stationary points:
\\
\\
\\
(1)~~$X= -1,~ Y= 0$\\
In this case, the corresponding eigenvalues are \\
$\mu_1 = 6 + 3/(-1 + \alpha) \ < 0$,~~ for $\alpha \ < 1$, \qquad
$\mu_2 = 3 + \sqrt{3/2}~ \lambda \ < 0$,~~ for $\lambda \ < - \sqrt{6}$
\\
The eigenvalues of this point show the negativity for $\alpha \ < 1$ and $\lambda \ < - \sqrt{6}$. Therefore, it is stable point.
\\
\\
\\
(2)~~$X= 0,~ Y= 0$\\
In this case, the corresponding eigenvalues are \\
$\mu_1 = 3/(2 -2 \alpha) \ < 0$,~~ for $\alpha \ > 1$, \qquad
$\mu_2 = -3 +  3/(2 -2 \alpha) \ < 0$,~~ for $1 \ < \alpha \ < 1/2$
\\
The negativity of the eigenvalues represents the stability. This point is stable for $\alpha \ > 1$.
\\
\\
\\
(3)~~$X= 1,~ Y= 0$\\
In this case, the corresponding eigenvalues are \\
$\mu_1 = 6 + 3/(-1 + \alpha) \ < 0$,~~ for $\alpha \ < 1$, \qquad
$\mu_2 = 3 - \sqrt{3/2}~ \lambda \ < 0$,~~ for $\lambda \ < \sqrt{6}$
\\
This point is stable for $\alpha \ < 1$ and $\lambda \ < \sqrt{6}$.
\\
\\
\\
(4)~~$X= \frac{\lambda}{\sqrt{6}},~ Y= \sqrt{1-\frac{\lambda^{2}}{6}}$\\
In this case, the corresponding eigenvalues are \\
$\mu_1 = (-6 + \lambda^2)/2 \ < 0$,~~ for $\lambda \ < \sqrt{6}$, \qquad
$\mu_2 = 3/(-1 + \alpha) + \lambda^2 \ < 0$,~~ for $\alpha \ < 1$,~ $\lambda \ < \sqrt{3/(-1 + \alpha)}$
\\
This point is stable under above given conditions.
\\
\\
\\
(5)~~$X= \frac{\sqrt{\frac{3}{2}}}{\lambda(1-\alpha)},~ Y= \frac{\sqrt{\frac{3}{2}-3\alpha}}{\lambda(\alpha-1)}$ \\
In this case, the corresponding eigenvalues are \\
\\
$\mu_1 = -\frac{3(\lambda^{2}(\alpha-1)^{2}(2\alpha-1)+\sqrt{\lambda^{2}(\alpha-1)^{3}(2\alpha-1)(24+\lambda^{2}(7+2\alpha)(\alpha-1})}{4\lambda^{2}(\alpha-1)^{3}} \ < 0$,~~ for $\alpha \ < -7/2,~ \lambda \ > 0$,
\\
\\
$\mu_2 = \frac{3(\lambda^{2}(\alpha-1)^{2}(2\alpha-1)+\sqrt{\lambda^{2}(\alpha-1)^{3}(2\alpha-1)(24+\lambda^{2}(7+2\alpha)(\alpha-1})}{4\lambda^{2}(\alpha-1)^{3}} \ < 0$,~~ for $\alpha \ < -7/2$,~ $\lambda \ > \sqrt{2/3}$
\\

This point has negative eigenvalues for $\alpha \ < -7/2$ and $\lambda \ > \sqrt{2/3}$. Therefore, it is stable point. We are interested in this point because it has both the parameters $\alpha$ and $\lambda$. We evolve the autonomous system (\ref{n18}) numerically for the parameter values $\alpha= -3.6$, $\lambda=0.94$, and $\alpha= -4.6$, $\lambda=0.94$, and the obtained results are shown in figure \ref{figint1}. For the chosen parameters the stable point behaves as an attractive node which is confirmed by the panels (a) and (b) of figure \ref{figint1}. The lower panels of figure \ref{figint1} show the scaling behaviour that gives acceleration. In addition we also calculate basic cosmological observables $W_{tot}$, $W_{\phi}$, $\Omega_{\phi}$ and obtained as $-$0.78, $-$1.06, 0.73 and $-$0.82, $-$1.36, 0.60 corresponding to $\alpha= -3.6$, $\lambda=0.94$, and $\alpha= -4.6$, $\lambda=0.94$, respectively. This point is summarized in table \ref{tabparm}. 
\subsection{Interacting model II}
\label{M2}
This model is specified by a coupling of the form
\begin{equation}
A=\beta\dot{\rho_{\phi}}
\end{equation}
For this coupling term we have
\begin{eqnarray} 
\label{Hd2}
\frac{\dot{H}}{H^2}&=&-\frac{3}{2} \left[1-Y^{2}+\frac{X^{2}(1-\beta)}{1+\beta} \right]\\ 
\label{n19}
\dfrac{dX}{dN}&=&-3X + \sqrt{\frac{3}{2}}\lambda Y^{2} +\frac{3}{2}X\left[1-Y^{2}+\frac{X^{2}(1-\beta)}{1+\beta} \right]\\ \nonumber
\dfrac{dY}{dN}&=&-\sqrt{\frac{3}{2}}\lambda XY +\frac{3}{2}Y\left[1-Y^{2}+\frac{X^{2}(1-\beta)}{1+\beta} \right]
\end{eqnarray}
Using equations (\ref{wtot}) and (\ref{Hd2}), the total equation of state for this model can be written as
\begin{eqnarray}
\label{wtot2}
W_{tot}&=&-1 + \left[1-Y^{2}+\frac{X^{2}(1-\beta)}{1+\beta} \right]\equiv W_{\phi} \Omega_{\phi}
\end{eqnarray}
\begin{figure*}
\begin{center}
$\begin{array}{c@{\hspace{1in}}c}
\epsfxsize=2.4in
\epsffile{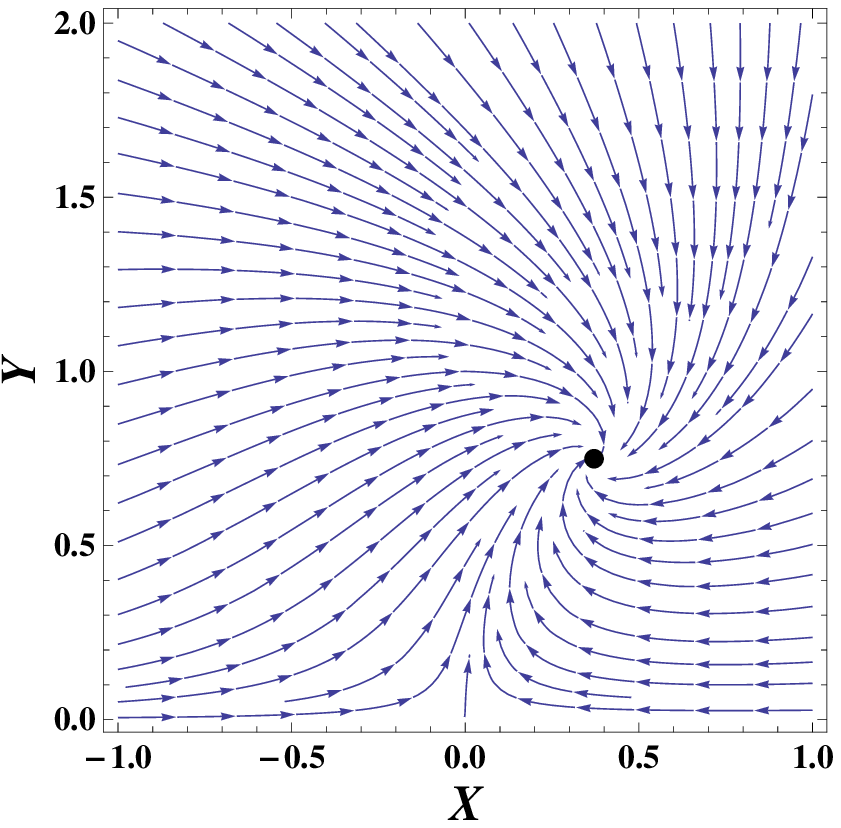} &
	\epsfxsize=2.4in
	\epsffile{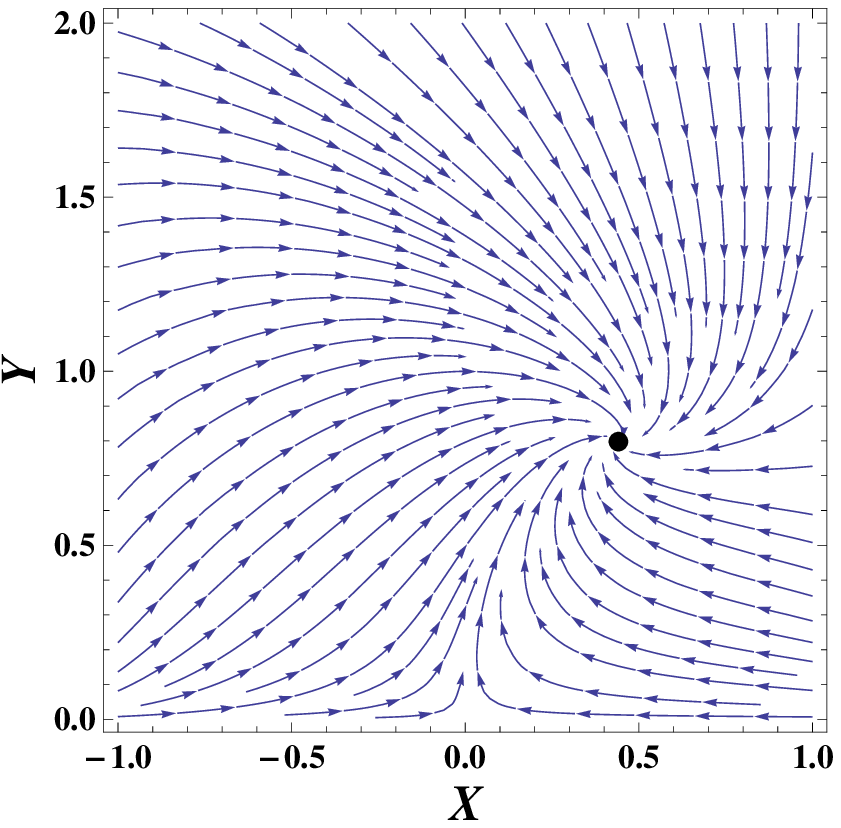} \\ [0.4cm]
\mbox{\bf (a)} & \mbox{\bf (b)}
\end{array}$
$\begin{array}{c@{\hspace{1in}}c}
\epsfxsize=2.4in
\epsffile{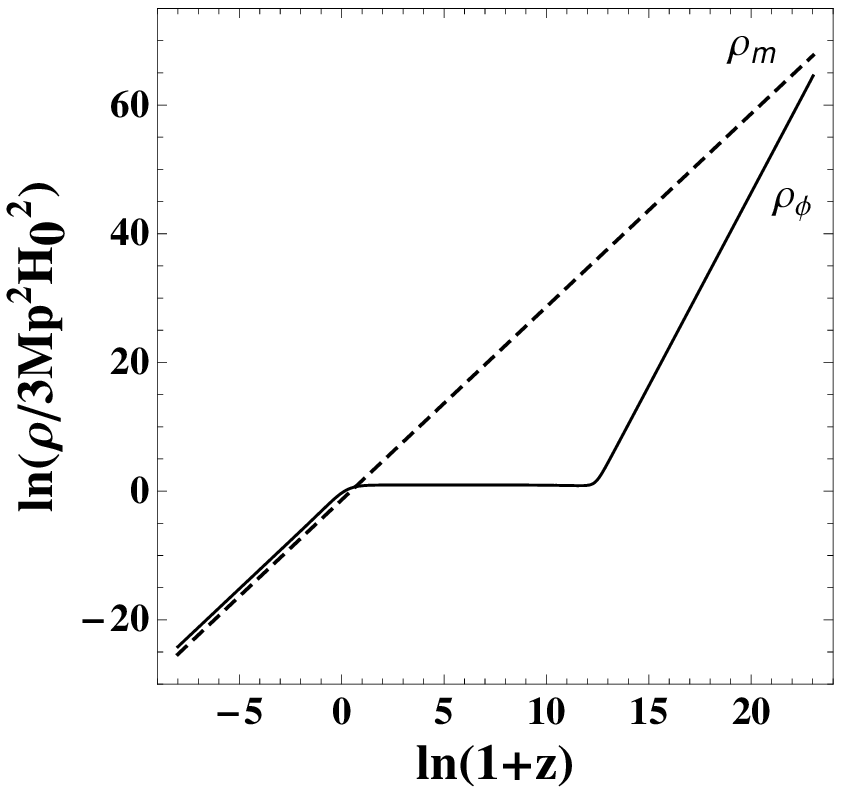} &
	\epsfxsize=2.4in
	\epsffile{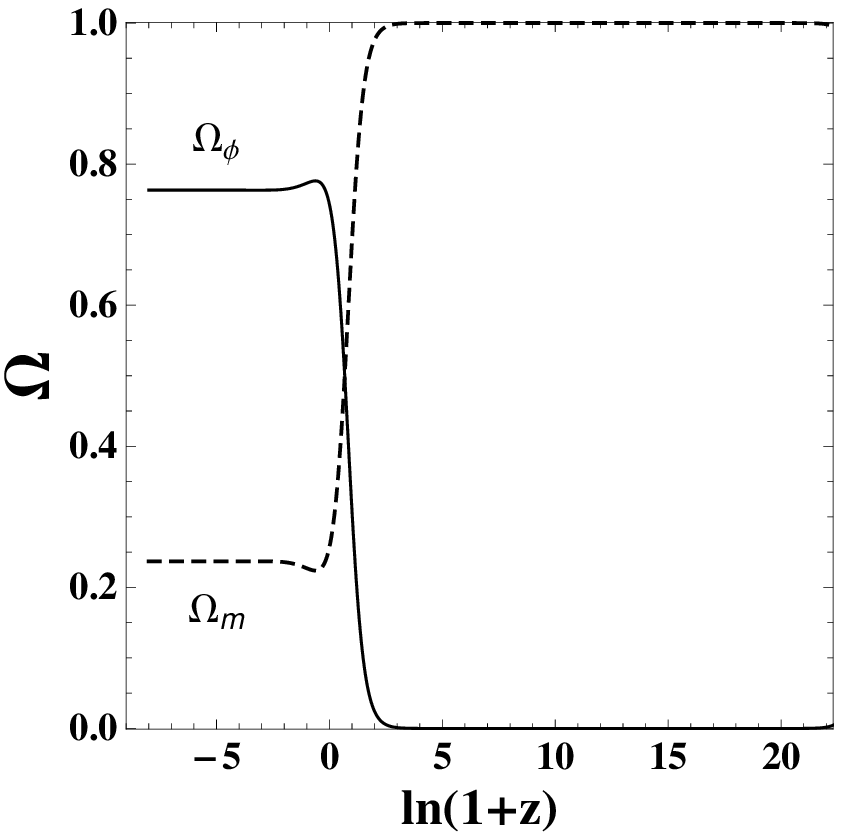} \\ [0.4cm]
\mbox{\bf (c)} & \mbox{\bf (d)}
\end{array}$
\end{center}
\caption{ \small The figure represents phase space trajectories, evolution of energy density and density parameter for interacting model II for the stable fixed point 3. The panel (a) corresponds to $\beta= 0.7$ and $\lambda= 1.3$, for these values of the parameters we obtain $\Omega_{\phi}=0.76$, $W_{tot}=-0.57$, $W_{\phi}=-0.75$ and an accelerating attractor solution. The panel (b) corresponds to $\beta= 0.3$ and $\lambda= 1.3$, correspondingly we get $\Omega_{\phi}=0.83$, $W_{tot}=-0.53$, $W_{\phi}=-0.64$ and an accelerated attractor solution. In both the panels, black dots designate attractor stable point, and the stable point behaves as an attractive focus under the chosen parameters. The panels (c) and (d)  are plotted the same values of the parameters as panel (a), and show scaling behaviours that gives an accelerated expansion.}
\label{figint2}
\end{figure*}
This interacting model has following stationary points:
\\
\\
\\
(1)~~$X= 0,~ Y= 0$\\
In this case, the corresponding eigenvalues are \\
$\mu_1 = -3/2$, \qquad
$\mu_2 = 3/2$
\\
This is unstable point because one of the eigenvalue is positive.
\\
\\
\\
(2)~~$X= \frac{\sqrt{-1-\beta}} {\sqrt{-1+\beta}}, ~Y= 0$\\
In this case, the corresponding eigenvalues are \\
$\mu_1 = 3$, \qquad
$\mu_2 = 3 - \frac{\sqrt{-3 (1+\beta)/2}~ \lambda} {\sqrt{-1+\beta}} < 0$,~~ for $\sqrt{-3 (1+\beta)/2}~ \lambda > 3(-1+\beta)$
\\
It is also unstable point because its one eigenvalue is positive.
\\
\\
\\
(3)~~$X= \frac{(3 + \lambda^2)(1 +\beta)-\delta} {2 \sqrt{6}~ \lambda}$,\qquad
$Y= \frac{\sqrt{(1-\beta^2)(6+9/ \lambda^2)-(1+\beta^2)\lambda^2+\delta(1+ \beta-3/\lambda^2+ 3 \beta / \lambda^2) }}  {2\sqrt{3}}$
\\
In this case, the corresponding eigenvalues are \\
\\
$\mu_1 =$ \noindent\({\frac{1}{8} \left(3 \left(-5+\lambda ^2\right)+3 \beta  \left(3+\lambda ^2\right)-3\text{  }\delta - \surd \left(\frac{2}{\lambda
^2}\left(216-63 \lambda ^2+\lambda ^6+\beta ^2 \left(-24+\lambda ^2\right) \left(3+\lambda ^2\right)^2-72 \delta -3 \lambda ^2 \delta -\lambda ^4
\delta +\right.\right.\right.}\\
{\left.\left.\left.\beta  \left(2 \lambda ^6+72 \delta -\lambda ^4 (18+\delta )+3 \lambda ^2 (30+7\delta  )\right)\right)\right)\right)}\)
\\
\\ 
$\mu_2 = $  \noindent\({\frac{1}{8} \left(3 \left(-5+\lambda ^2\right)+3 \beta  \left(3+\lambda ^2\right)-3\text{  }\delta + \surd \left(\frac{2}{\lambda
^2}\left(216-63 \lambda ^2+\lambda ^6+\beta ^2 \left(-24+\lambda ^2\right) \left(3+\lambda ^2\right)^2-72 \delta -3 \lambda ^2 \delta -\lambda ^4
\delta +\right.\right.\right.}\\
{\left.\left.\left.\beta  \left(2 \lambda ^6+72 \delta -\lambda ^4 (18+\delta )+3 \lambda ^2 (30+7\delta  )\right)\right)\right)\right)}\),
\\
\\
where
\begin{equation}
\label{delta}
\delta= \sqrt{(1+\beta)\left((-3+\lambda ^2)^2+\beta (3+\lambda ^2)^2   \right)}
\end{equation}
\\
Above eigenvalues are negative for $-\infty \leq \beta \leq \infty$ and $-\infty \leq \lambda \leq \infty$ (but $\lambda \neq 0$).
\\
\\

The negativity of the eigenvalues exhibits the stability. This point is stable for all the values of $\beta$ and $\lambda$ provided that $\lambda \neq 0$. We elaborate the autonomous system (\ref{n19}) numerically for the parameter choices $\beta=0.7$, $\lambda=1.3$, and $\beta=0.3$, $\lambda=1.3$, and obtain $W_{tot}$, $W_{\phi}$ and $\Omega_{\phi}$ as $-$0.57, $-$0.75, 0.76 and $-$0.53, $-$0.64, 0.83, respectively. The phase space trajectories of this stable point are shown in panels (a) and (b) of figure \ref{figint2}, and the point behaves as an attractive focus. The lower panels of figure \ref{figint2} exhibit the scaling behaviour that gives late time acceleration. The results of the stable point are abbreviated in table \ref{tabparm}.
\begin{figure*}
\begin{center}
$\begin{array}{c@{\hspace{1in}}c}
\epsfxsize=2.4in
\epsffile{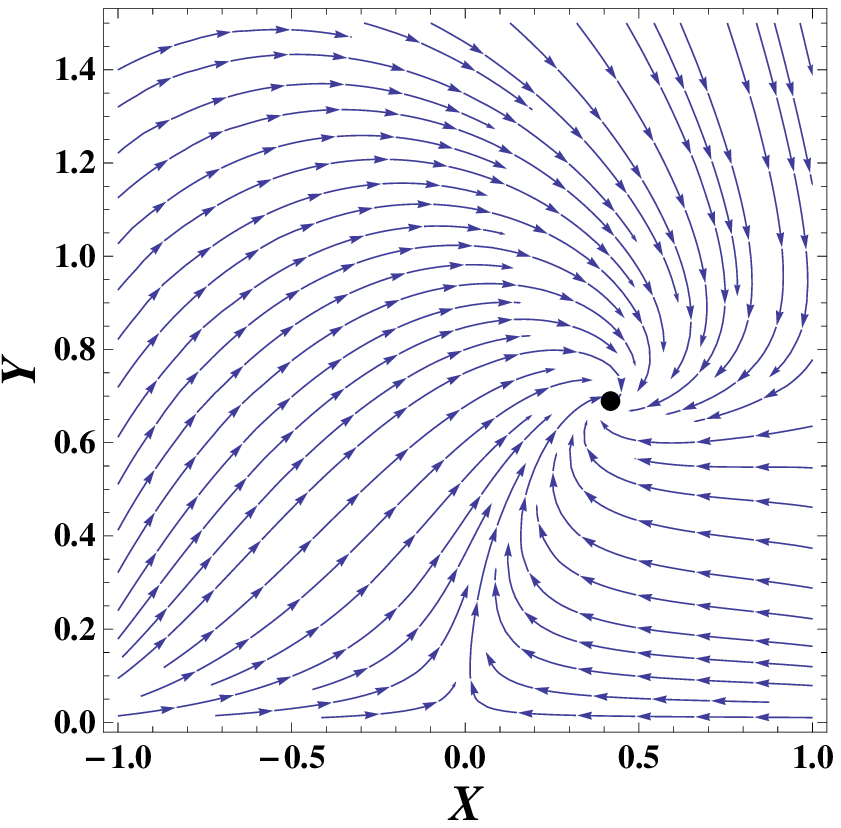} &
	\epsfxsize=2.4in
	\epsffile{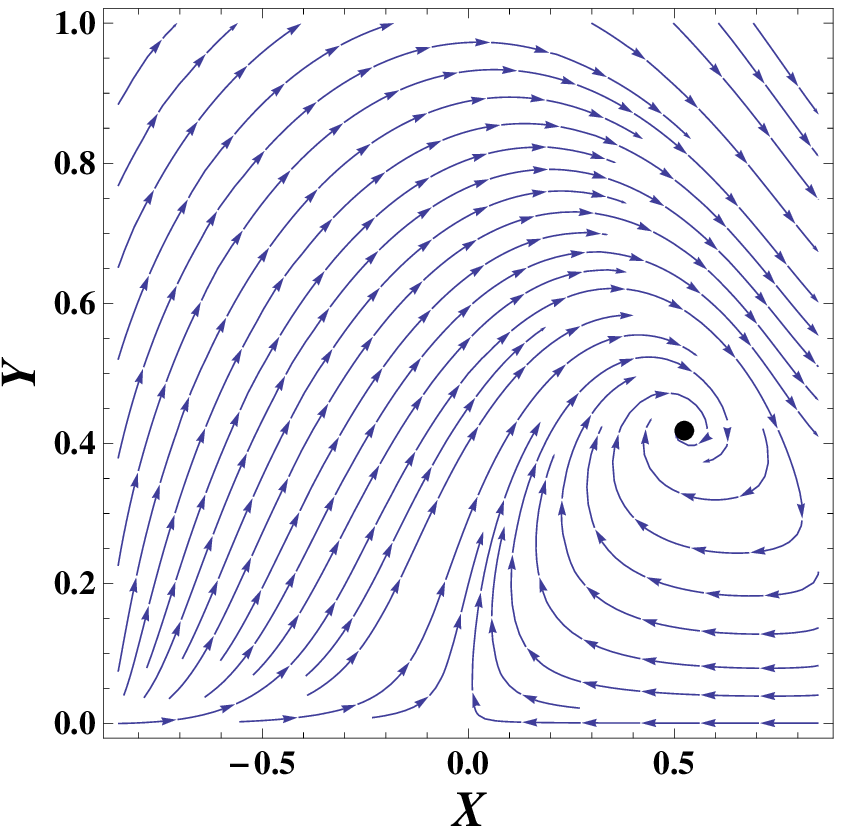} \\ [0.4cm]
\mbox{\bf (a)} & \mbox{\bf (b)}
\end{array}$
$\begin{array}{c@{\hspace{1in}}c}
\epsfxsize=2.4in
\epsffile{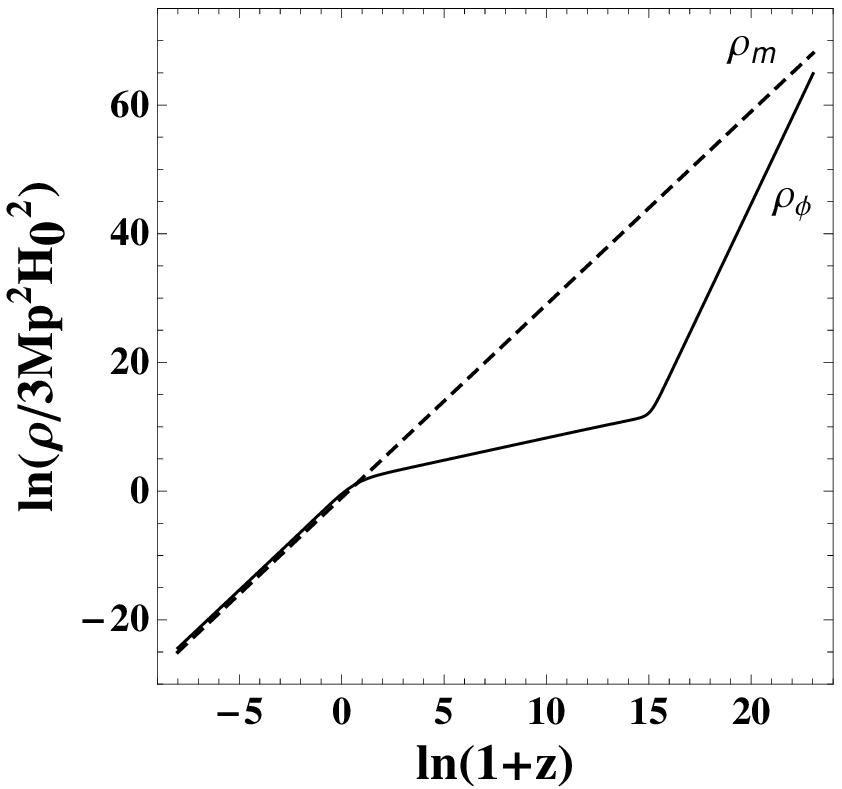} &
	\epsfxsize=2.4in
	\epsffile{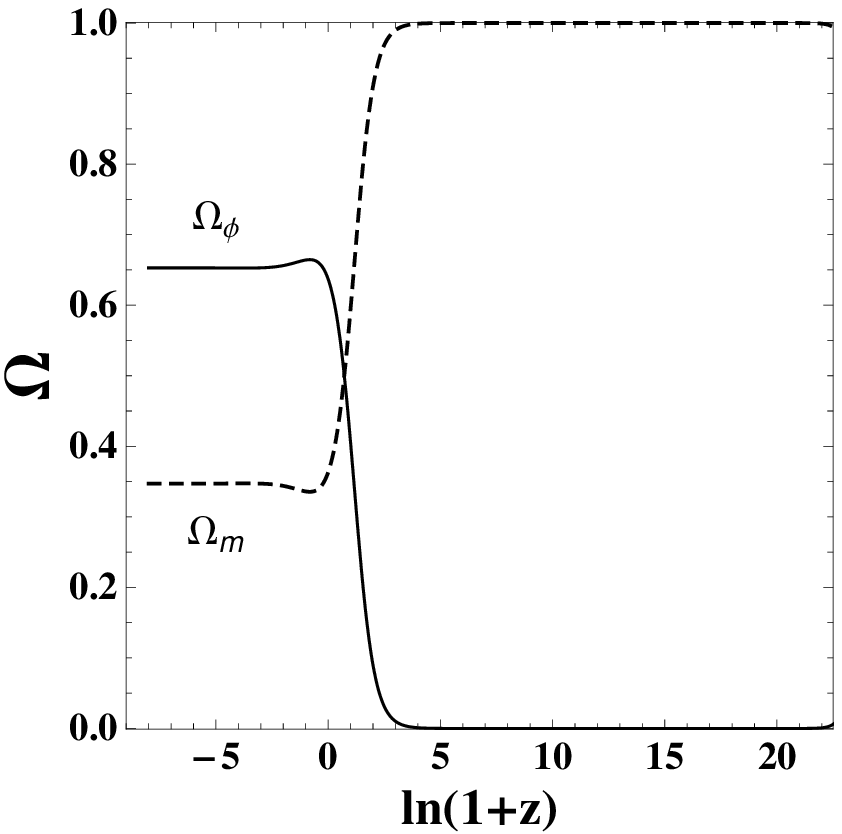} \\ [0.4cm]
\mbox{\bf (c)} & \mbox{\bf (d)}
\end{array}$
\end{center}
\caption{ \small The figure displays phase space trajectories, evolution of energy density and density parameter for interacting model III for the stable fixed point 3. The panel (a) corresponds to $\sigma= -0.3$ and $\lambda= 1.57$, for these values of the parameters we get $\Omega_{\phi}=0.65$, $W_{tot}=-0.46$, $W_{\phi}=-0.71$ and an accelerating attractor solution. The panel (b) corresponds to $\sigma= 0.1$ and $\lambda= 2.85$, and correspondingly we obtain $\Omega_{\phi}=0.45$, $W_{tot}=0.22$, $W_{\phi}=0.49$, since total equation of state is positive therefore, attracting solution but not accelerating. In both the panels, black dots represent attractor stable point, and the stable point acts as an attractive focus under the chosen parameters. For panels (c) and (d), we use same values of the parameters as panel (a). Both panels are showing attractor behaviour that corresponds to scaling solutions.}
\label{figint3}
\end{figure*}
\subsection{Interacting model III}
\label{M3}
In this model we consider the coupling form as a linear combination of $\dot{\rho_{m}}$ and $\dot{\rho_{\phi}}$ as
\begin{equation}
A=\sigma (\dot{\rho_{m}}+\dot{\rho_{\phi}})
\end{equation}
For this interaction form we have
\begin{eqnarray}
\label{HdIII}
\frac{\dot{H}}{H^2}&=&-\frac{3}{2}\left[\frac{1+X^{2}-Y^{2}}{1-\sigma}\right]\\
\label{auto3}
\dfrac{dX}{dN}&=&-3X + \sqrt{\frac{3}{2}}\lambda Y^{2} +\frac{3}{2}X\left[\frac{1+X^{2}-Y^{2}}{1-\sigma}\right]\\
\dfrac{dY}{dN}&=&-\sqrt{\frac{3}{2}}\lambda XY +\frac{3}{2}Y\left[\frac{1+X^{2}-Y^{2}}{1-\sigma}\right] \nonumber
\end{eqnarray}
The total equation of state for this model can be obtained by using equations (\ref{wtot}) and (\ref{HdIII}) as
\\
\begin{eqnarray}
\label{wtot3}
W_{tot}&=&-1 + \left[\frac{1+X^{2}-Y^{2}}{1-\sigma}\right]\equiv W_{\phi} \Omega_{\phi}
\end{eqnarray}
\\
This model has following stationary points:
\\
\\
\\
(1)~~$X= 0,~ Y= 0$\\
In this case, the corresponding eigenvalues are \\
$\mu_1 = -\frac{3}{2(-1+\sigma)} < 0$, for $\sigma > 1$ \qquad
$\mu_2 = \frac{3-6\sigma}{2(-1+\sigma)} < 0$, for $\sigma > 1/2$
\\
This point is stable for $\sigma > 1$.
\\
\\
\\
\\
(2)~~$X= \sqrt{1-2 \sigma},~ Y= 0$\\
In this case, the corresponding eigenvalues are \\
$\mu_1 = 6+\frac{3}{-1+\sigma}< 0$, for $\sigma < 1/2$\qquad
$\mu_2 = 3 - \sqrt{3/2-3\sigma }~\lambda < 0$,~~ for $\sqrt{3/2-3\sigma }~\lambda > 3$
\\
This point is also stable for the above given conditions.
\\
\\
\\
\\
(3)~~$X= $ \noindent\({\frac{9-(-1+\sigma )^2 \lambda ^4+\eta }{2 \sqrt{6} \lambda  \left(3+(-1+\sigma ) \lambda ^2\right)}}\),\qquad
$Y=$ \noindent\({\frac{\sqrt{9+6 \lambda ^2-(-1+\sigma )^2 \lambda ^4+\eta }}{2 \sqrt{3}\lambda }}\)
\\
In this case, the corresponding eigenvalues are \\
\\
\\
\\
$\mu_1 =$ \noindent\({\frac{1}{8} \left(-3+\frac{12}{-1+\sigma }-3 (-1+\sigma ) \lambda ^2+\frac{3 \eta }{3+(-1+\sigma ) \lambda ^2}- \surd 2\left(\frac{9
(-7+\sigma  (2+13 \sigma ))}{(-1+\sigma )^2}-\frac{216}{(-1+\sigma ) \lambda ^2}+6 \sigma  \lambda ^2+(-1+\sigma )^2 \lambda ^4- \eta \right.\right.}\\
{\left.\left.-\frac{24\text{  }\eta }{(-1+\sigma ) \lambda ^2}-\frac{30\text{  }\eta }{(-1+\sigma ) \left(3+(-1+\sigma ) \lambda ^2\right)}+\frac{18
\sigma \text{  }\eta }{(-1+\sigma ) \left(3+(-1+\sigma ) \lambda ^2\right)}\right)\right)}\\
{}\)
\\
\\ 
\\
\\
$\mu_2 = $ \noindent\({\frac{1}{8} \left(-3+\frac{12}{-1+\sigma }-3 (-1+\sigma ) \lambda ^2+\frac{3 \eta }{3+(-1+\sigma ) \lambda ^2}+ \surd 2\left(\frac{9
(-7+\sigma  (2+13 \sigma ))}{(-1+\sigma )^2}-\frac{216}{(-1+\sigma ) \lambda ^2}+6 \sigma  \lambda ^2+(-1+\sigma )^2 \lambda ^4- \eta \right.\right.}\\
{\left.\left.-\frac{24\text{  }\eta }{(-1+\sigma ) \lambda ^2}-\frac{30\text{  }\eta }{(-1+\sigma ) \left(3+(-1+\sigma ) \lambda ^2\right)}+\frac{18
\sigma \text{  }\eta }{(-1+\sigma ) \left(3+(-1+\sigma ) \lambda ^2\right)}\right)\right)}\\
{}\)
\\
\\
\\
where
\begin{equation}
\label{eta}
\eta = \sqrt{\left(3+(-1+\sigma ) \lambda ^2\right)^2 \left(9-6 (1+\sigma ) \lambda ^2+(-1+\sigma )^2 \lambda ^4\right)}
\end{equation}
\\
Above eigenvalues are negative for $\sigma < 0.2$ provided that $\sigma \neq  -1$ and $ \lambda \leq \sqrt{\frac{6(1+\sigma)}{(-1+\sigma)^2}}$. 
\\
\\

This point has both the parameters $\sigma$, $\lambda$ and shows the stability for the choices of the parameters $\sigma < 0.2$ ($\sigma \neq  -1$) and $ \lambda \leq \sqrt{\frac{6(1+\sigma)}{(-1+\sigma)^2}}$. We numerically elaborate the autonomous system (\ref{auto3}) for the parameter choices $\sigma=-0.3$, $\lambda=1.57$, and $\sigma=0.1$, $\lambda=2.85$, and obtain $W_{tot}$, $W_{\phi}$, $\Omega_{\phi}$ as $-$0.46, $-$0.71, 0.65 and 0.22, 0.49, 0.45, respectively. We do not find accelerating solution corresponding to $\sigma=0.1$ as it has positive equation of state. Also, we check numerically for all positive values of $\sigma$ ($0< \sigma <0.2$) and do not find accelerating phase. The results of this stable point are concised in table \ref{tabparm}. The phase portraits of the point are displayed in panels (a) and (b) of figure \ref{figint3}, and the point acts as an attractive focus. The lower panels of figure \ref{figint3} exhibit the scaling behaviour that provides late time acceleration.
\begin{table}
\caption{We present the stable points 5, 3 and 3 for interacting models I, II and III respectively, these stable points have two parameters, namely, $\alpha, \lambda$ and $\beta, \lambda$ and $\sigma, \lambda$. We also introduce the expressions of $\Omega_{\phi}$, $W_{tot}$ and the condition of acceleration in terms of the parameters. Two numerical choices of the parameters are shown for each model. The symbols $\delta$ and $\eta$ are used for shorting the expressions and given by equations (\ref{delta}) and (\ref{eta}) respectively.}
\begin{center}
\label{tabparm}
\begin{tabular}{l c c c c c c r} 
\hline\hline 
\\
Model &$X$  &$Y$ &Stable for &$\Omega_{\phi}$ &$W_{tot}$ & $W_{\phi}=$ & Acceleration\\
& & & & & & $\frac{W_{tot}}{\Omega_{\phi}}$ &\\
\\
\hline
\\
I  &  $\frac{\sqrt{3/2}}{\lambda(1-\alpha)}$ &   $\frac{\sqrt{3/2-3\alpha}}{\lambda(\alpha-1)}$ &  $\alpha< -7/2$,  &  $\frac{3}{\lambda^2(1-\alpha)}$  & $\frac{\alpha}{1-\alpha}$ & &for $\alpha< -1/2$\\
&&&$\lambda > \sqrt{2/3}$&&&&\\
\\
&&&$\alpha=-3.6$& 0.73 & $-0.78$ & $-1.06$ &Yes\\
&&&$\lambda=0.94$&&&\\
\\
&&&$\alpha=-4.6$& 0.60 & $-0.82$ & $-1.36$ &Yes\\
&&&$\lambda=0.94$&&&\\
\\
\hline
\\ 
II& $\frac{1} {2 \sqrt{6}~ \lambda}$ & $\frac{1}{2\sqrt{3}} \surd{[(1-\beta^2)(6+9/ \lambda^2)}$ & for all values of& $\frac{(3+\lambda^2)(1+\beta)-\delta}{2\lambda^2}$ &$\frac{\lambda^2-3+\beta(3+\lambda^2)-\delta}{6}$&&$\lambda^2+\beta(3+\lambda^2)$\\
& $[(3 + \lambda^2)$ &$-(1+\beta^2)\lambda^2 +\delta(1+ \beta$& $\beta$ and $\lambda$ provided&&&&$-\delta <  1$\\
&$(1 +\beta)-\delta]$&$-3/\lambda^2+ 3 \beta / \lambda^2)]$& that $\lambda \neq 0$
\\
\\
&&& $\beta=0.7$& 0.76& $-$0.57& $-$0.75&Yes\\
&&&$\lambda=1.3$&&&&\\
\\
\\
&&& $\beta=0.3$& 0.83& $-$0.53&$-$0.64& Yes\\
&&&$\lambda=1.3$&&&&\\
\\
\hline
\\
III& \({\frac{9-(1-\sigma )^2 \lambda ^4+\eta }{2 \sqrt{6} \lambda  \left(3-(1-\sigma ) \lambda ^2\right)}}\) &\({\frac{\sqrt{9+6 \lambda ^2-(1-\sigma )^2 \lambda ^4+\eta }}{2 \sqrt{3}\lambda }}\)& $\sigma < 0.2 \neq -1$& $\frac{9-(1-\sigma)^2 \lambda^4 +\eta}{6\lambda^2-2(1-\sigma) \lambda^4}$ & $ \frac{1}{18-6(1-\sigma) \lambda^2}$ & &$27-3(1-\sigma) \lambda^2$ \\
&&&$\lambda \leq \frac{\sqrt{6(1+\sigma)}}{1-\sigma}$&& $[(1-\sigma) \lambda^2$ & &$[4-(1-\sigma) \lambda^2]$\\
&&&&&$(6-(1-\sigma) \lambda^2 )$&&$-3\eta < 0$\\
&&&&&$+ \eta-9]$&&
\\
\\
&&& $\sigma=-0.3$& 0.65& $-$0.46&$-$0.71 &Yes\\
&&&$\lambda=1.57$&&&&\\
\\
\\
&&& $\sigma=0.1$& 0.45& 0.22& 0.49&No\\
&&&$\lambda=2.85$&&&&\\
\\
\hline\hline 
\end{tabular}    
\end{center}
\end{table}
\section{Conclusions}
\label{conc}
We studied interaction of quintessence with dark matter in spatially flat  Universe. In the absence of fundamental theory of specific interaction in the dark sector, the choice of interaction strength in the conservation of energy equations was phenomenological and heuristic. In this paper we considered three phenomenological interacting quintessence cosmological models as $\alpha \dot{\rho_m}$, $\beta \dot{\rho_{\phi}}$, and $\sigma(\dot{\rho_{m}}+ \dot{\rho_{\phi}})$. Our primary object was to inspect whether there exist late time accelerated scaling attractor  having $\Omega_{DE}/\Omega_{DM}$ of the order one. We studied dynamical behaviour and phase space analysis of the models under consideration. We focussed on the stable points which could give rise to scaling attractors. In all the models we obtained fundamental cosmological observables like $\Omega_{\phi}$, $W_{tot}$, $W_{\phi}$ corresponding to two numerical choices of the parameters. For the interacting model I we found that the fixed point 5 is stable for $\alpha < -7/2$ and $\lambda > \sqrt{2/3}$. The phase space trajectories, evolution of energy density and density parameter for different numerical choices of the parameters are shown in figure \ref{figint1}. The fixed point 5, in this case, corresponds to accelerated scaling attractor  with $\Omega_{DE}/\Omega_{DM}= O(1)$. The point 3 of interacting model II shows stability for all values of $\beta$ and $\lambda$ provided that $\lambda \neq 0$. The phase portraits, evolution of energy density and density parameter for different numerical values of the parameters are displayed in figure \ref{figint2}. Clearly, this is scaling solution with the required property. For interacting model III we noticed that the fixed point 3 is stable for $\sigma < 0.2 \neq -1$ and $\lambda \leq \sqrt{6(1+\sigma)}/(1- \sigma)$. Figure \ref{figint3} shows the phase space trajectories, evolution of energy density and density parameter for different values of the parameters. Our analysis shows that accelerating attractor, in this case, exists only for negative values of $\sigma$. The lower panels of figure \ref{figint3} shows accelerated attractor solution for $\sigma= -0.3$ and $\lambda=1.57$.
For all the models, we obtained late time accelerated scaling attractor having $\Omega_{DE}/\Omega_{DM}= O(1)$. Therefore all the models considered in this paper are viable to solve the coincidence problem.
 
\begin{acknowledgments}
M.S. thanks M. Sami for his useful comments and suggestions. S.D.P. acknowledges IUCAA, Pune and CTS, IIT Kharagpur for their hospitalities under visiting programme and  thanks V. Sahni for his constant help and support. The work by M.Kh. on initial cosmological conditions was supported by the Ministry of Education and Science of Russian Federation, project 3.472.2014/K  and his work on the forms of dark matter was supported by grant RFBR 14-22-03048. M.M.V. thanks Edward W. Kolb for discussions and hospitality at the Kavli Institute of Cosmological Physics, the University of Chicago. 
\end{acknowledgments}

\end{document}